\documentclass[runningheads]{llncs}
\usepackage[T1]{fontenc}
\usepackage{graphicx}
\usepackage{amsfonts}
\usepackage{array}
\usepackage{booktabs}       

\usepackage{ amssymb }

\newcolumntype{P}[1]{>{\centering\arraybackslash}p{#1}}

\begin{document}
\title{MeshBrush: Painting the Anatomical Mesh with Neural Stylization for Endoscopy}
\titlerunning{MeshBrush: Neural Mesh Stylization for Endoscopy}
%
\author{John J. Han\inst{1} 
\and Ayberk Acar\inst{1} 
\and Nicholas Kavoussi\inst{2} 
\and Jie Ying Wu\inst{1}
}

\authorrunning{J. J. Han et al.}

\institute{Vanderbilt University, Nashville TN, USA \\ \email{\{john.j.han, ayberk.acar, jieying.wu\}@vanderbilt.edu} \and Vanderbilt University Medical Center, Nashville TN, USA\\ \email{nicholas.l.kavoussi@vumc.org}
}
\maketitle 
\begin{abstract}
Style transfer is a promising approach to close the sim-to-real gap in medical endoscopy. Rendering synthetic endoscopic videos by traversing pre-operative scans (such as MRI or CT) can generate structurally accurate simulations as well as ground truth camera poses and depth maps. Although image-to-image (I2I) translation models such as CycleGAN can imitate realistic endoscopic images from these simulations, they are unsuitable for video-to-video synthesis due  to the lack of temporal consistency, resulting in artifacts between frames. We propose \textit{MeshBrush}, a neural mesh stylization method to synthesize temporally consistent videos with differentiable rendering. \textit{MeshBrush} uses the underlying geometry of patient imaging data while leveraging existing I2I methods. With learned per-vertex textures, the stylized mesh guarantees consistency while producing high-fidelity outputs. We demonstrate that mesh stylization is a promising approach for creating realistic simulations for downstream tasks such as training networks and preoperative planning. Although our method is tested and designed for ureteroscopy, its components are transferable to general endoscopic and laparoscopic procedures. The code will be made public on GitHub\footnote{https://github.com/juseonghan/MeshBrush}.

\end{abstract}

\section{Introduction}

Simulated environments have a wide array of applications in endoscopy, such as surgical training \cite{khan2018virtual}, medical education \cite{khan2019sim}, and patient-specific preoperative planning \cite{SHINOMIYA201813}. Furthermore, these environments have been used in ground truth generation of computer vision tasks such as endoscopic depth estimation~\cite{rau2019implicit}. However, current simulated environments (e.g. renderings from a patient CT model) lack realism in depicting endoscopic surgical anatomy which limits clinical adoption~\cite{cardoso2023exploring}. To mitigate these challenges, leveraging real endoscopic video data to create temporally and spatially consistent videos could facilitate the development of accurate simulators and improve their practicality.

\begin{figure}
\includegraphics[width=\textwidth]{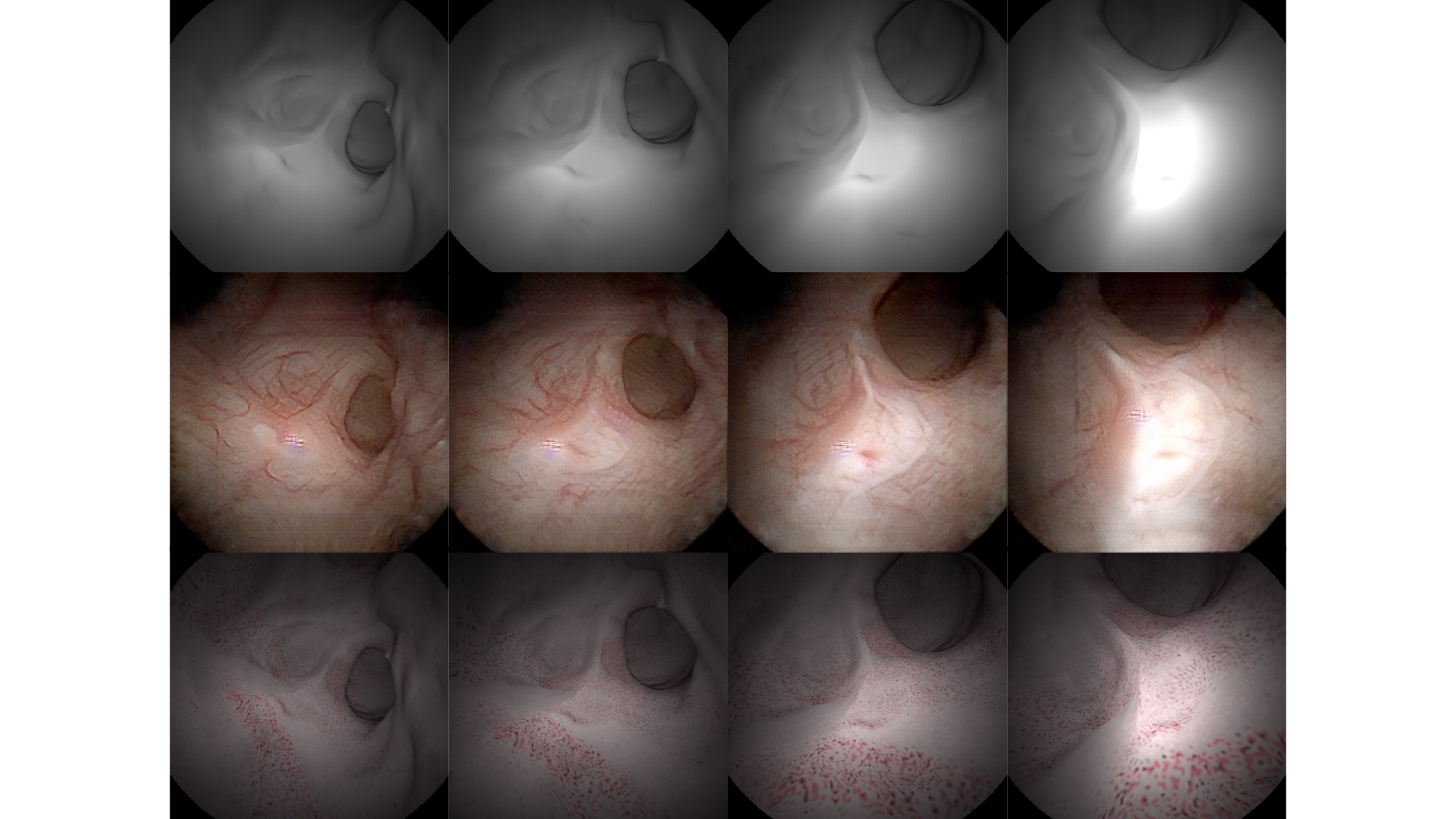}
\caption{Sample camera trajectory inside the mesh. From top to bottom: original rendered, I2I style transfers~\cite{pfeiffer2019generating}, and \textit{MeshBrush}. Note inconsistency from frame to frame in the I2I method while \textit{MeshBrush} maintains consistent features.}\label{demo}

\end{figure}

Image-to-image (I2I) translation, or style transfer, methods are promising approaches to synthesizing realistic endoscopy images from synthetic data such as CT-renders~\cite{lu2023assist} or to generating ground truth in monocular depth estimation for endoscopy~\cite{st_depth_tong}. For example, Tong et al. use it to translate real endoscopy images to rendered images and train a depth network~\cite{st_depth_tong}.
However, these approaches are unsuitable for video-to-video synthesis due to the lack of temporal consistency; synthesized features are not enforced to ``persist'' between frames, resulting in flickering and artifacts in the translated video sequence (shown in Fig.~\ref{demo}). As a result, I2I translation is not viable in algorithms with consistency requirements such as vision-based 3D reconstruction and localization. Methods that enforce consistency by either incorporating landmark detection between frames~\cite{Sharan_2022} or using optical flow~\cite{frischtemporal} produce imperfect results; optimizing a 2D task such as landmark detection and flow cannot guarantee temporal and long-term consistency, an inherently 3D property. We propose a novel approach for the generation of realistic and consistent endoscopic views via mesh stylization. 

To the best of our knowledge, our work is the first in 3D endoscopic mesh stylization to synthesize consistently style-transferred videos from rendered data. Because mesh textures are stationary, \textit{MeshBrush} guarantees consistency regardless of time and view. Leveraging existing I2I style transfer modules, our self-supervised network simply requires an input mesh of the patient anatomy and a pretrained style transfer module. Our lightweight model produces high-resolution spatial textures on the patient mesh to create realistic endoscopic sequences with temporal and global consistency using a view-dependent heatmap loss. Finally, we demonstrate that temporally consistent style transfer is effective in vision algorithms such as feature matching and Structure-from-Motion. 


\section{Related Work}
\textbf{Style Transfer in Medical Scenes.} Style transfer has been employed for a number of applications in endoscopic and laparoscopic scenes, such as for enhanced simulation and realistic renderings~\cite{lu2023assist}, specular light removal~\cite{funke2018generative}, smoke removal~\cite{venkatesh2020unsupervised}, and depth estimation~\cite{st_depth_tong}. CycleGAN~\cite{zhu2017unpaired} is commonly employed for image translation due to its ability to train with unpaired images from two domains. However, these methods do not ensure temporal or global consistency; each frame is processed independently with no propagating details between frames.

Few works address temporal and long-term consistent style transfer in medical scenes. Frisch et al. use optical flow (RAFT \cite{teed2020raft}) to warp views and impose a time-invariant structural similarity loss in cataract surgical video~\cite{frisch2023temporally}. However, as flow estimates are imperfect, the results are suboptimal. Engelhardt et al. extend CycleGAN~\cite{zhu2017unpaired} to process multiple consecutive frames simultaneously for multi-frame consistency in phantom surgical videos~\cite{engelhardt2018improving}, which cannot enforce long-term consistency, e.g. when the endoscope returns to a location after a certain duration. Rivoir et al. use the underlying 3D model to impose similar textures via a view consistency loss and learnable neural textures to ensure global consistency in simulated surgical scenes~\cite{rivoir2021longterm}. Although this method demonstrates remarkable performance, its applications are solely for laparoscopy which typically has simpler geometric structures than endoscopic anatomies. Mesh stylization ensures consistency regardless of time and viewpoint for complex 3D geometries such as simulated endoscopic environments. 



\textbf{Mesh Stylization and Consistent Video Style Transfer}. Stylizing a mesh from a given style is an established problem in the vision and graphics community. For instance, Ma et al. focus on altering the style of a mesh from a user-defined text prompt~\cite{ma2023xmesh}. Other stylization approaches use a variety of style inputs such as other meshes~\cite{KANG2023101198}, image-text pairs~\cite{Canfes_2023_WACV}, and image exemplars~\cite{Hollein_2022_CVPR}. 

Similarly, video style transfer is a well-known task in the general vision community. In addition to aforementioned works that use optical flow, recent work leverages further spatial control signals (e.g. depth maps) to account for noisy flow \cite{liang2023flowvid}. Chu et al. \cite{chu2023video} translate videos from synthetic 3D data, where ground truth optical flow and depth can be computed; however, this approach is text-guided using a ControlNet \cite{zhang2023adding}, making it unsuitable for endoscopic applications. 

\section{Methods}
\begin{figure}
\includegraphics[width=\textwidth]{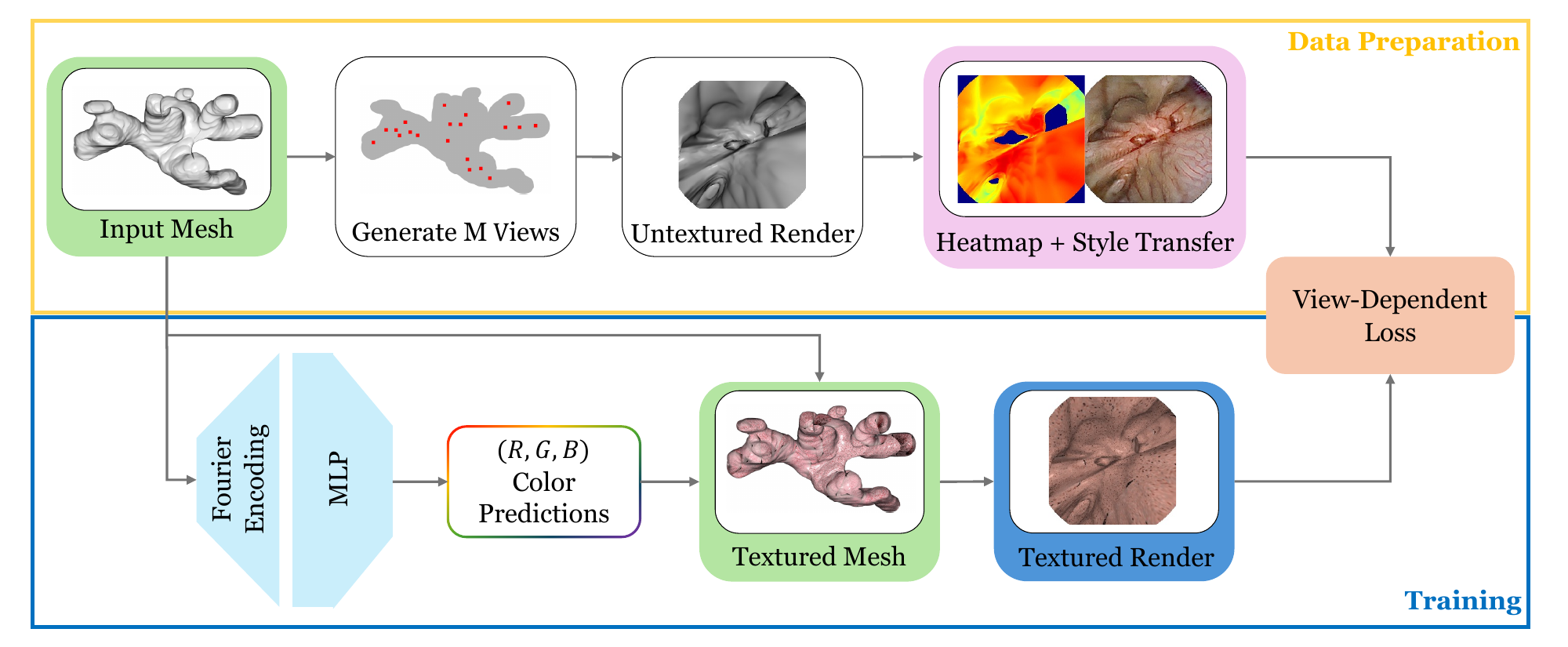}
\caption{An overview of our method. A skeletonization method \cite{skeletor} generates camera views inside the anatomy. At each iteration, the model re-renders from each camera pose via differential rendering and its style-transferred image supervises the vertex textures to be updated via the view-dependent heatmap loss.} \label{workflow}
\end{figure}

\textbf{Preprocessing and Data Preparation.} Fig.~\ref{workflow} shows the overview of \textit{MeshBrush}. First, we segment a mesh from a CT scan with a medical visualization toolkit\footnote{https://www.slicer.org/} and increase the vertex resolution by subdividing each triangle. This mesh is the structure upon which we will impose color information. We use an existing I2I method on specific camera renders and project these style transfer features onto the mesh. To generate meaningful mesh renders, we extract a mesh skeleton via~\cite{skeletor}, from which we generate $M$ camera positions with the farthest point down sampling. We set the camera view direction along its line segment of the skeleton to determine the $SE(3)$ camera poses $\{T_i\}_{i=1}^M$. 

We generate supervisory I2I renderings as data preparation. From each camera pose $T_i$, we render a synthetic, unrealistic image $A_i$. Using a pre-trained image-to-image style transfer module~\cite{pfeiffer2019generating}, we translate $A_i$ into a realistic endoscopic view $B_i$. Our modularized approach allows our framework to be used for various medical domains as users can change the I2I style transfer method to match each domain. For each rendered view, we also compute a heatmap $H_i$, explained in the next section, which encodes the 3D scene geometry relative to the camera. We compile a set of $(T_i, B_i, H_i)$ pairs for all $M$ views to use as ground truth supervision during the stylization process. 

\textbf{Training}. During training, our goal is to predict realistic endoscopic color values for each vertex of the mesh. We use a multilayer perception (MLP) network to regress RGB values; however, MLPs are not proficient in learning high-frequency associations such as texture variations. To overcome this issue, we first encode the vertex coordinates $V \in \mathbb{R}^{N \times 3}$ of the mesh to a Fourier feature encoding~\cite{tancik2020fourier}, i.e. $\gamma(V)=(\cos(2\pi GV), \sin(2\pi GV))$, where $G$ is a matrix with elements drawn from a zero-mean Gaussian distribution. The encoded vertices $\gamma(V)$ are passed into the MLP to generate texture predictions, which are applied onto the mesh. We then re-render a new set of images $\hat{A}_i$ from camera views $T_i$ from the data preparation phase. We supervise $\hat{A}_i$ with its corresponding style transferred image $B_i$ via the view-dependent loss, expounded in the next section. The MLP receives gradient updates from the error of $\hat{A}_i$ and $B_i$ via differentiable rendering. We repeat this over all $M$ camera views to complete one epoch of optimization. Over multiple epochs, the error between $\hat{A}_i$ and $B_i$ is minimized through training, which causes the mesh textures to mimic the style-transferred images.

Using a neural network to predict textures rather than projecting the style transfer RGB values to the mesh has two important advantages. First, vertices are much spatially finer than pixels, which enables the mesh to learn more realistic textures at a higher resolution. Second, projecting RGB values from $B_i$ introduces quilting, where camera poses with overlapping image regions introduce sharp texture discontinuities due to inconsistency in the transferred styles.

\textbf{View-dependent Heatmap Loss}. Vertices that have limited visibility or resolution from the perspective of the camera are often poorly textured in $B_i$, e.g. vertex normals perpendicular to the view direction or vertices far away from the camera. As a result, we devise a strategy for weighing certain pixels based on 1) the angle between a visible vertex normal and camera viewing direction, and 2) the distance of the visible vertices from the camera.

We omit the subscript $i$ in this section for the sake of clarity. Given a rendered image $A$, we calculate a normal map which contains the vertex normals at each pixel of the rendered image. By computing the dot product between the normal map and the camera viewing direction, we can calculate the cosine of the angle between each pixel's 3D point and the camera orientation, creating the view-orientation heatmap $\theta$. Similarily, the view-distance heatmap $D$ can be calculated by the $L_2$ distance between the camera position and 3D points in the rendered image. After scaling $D$ to unit, the heatmaps are aggregated to create the final heatmap $H \in \left[0,1\right]$. 


\begin{equation}
H = \frac{1}{2}\left( \frac{1 - \theta}{2} + D \right)
\end{equation}

The final heatmap is generated by forcing pixels with view-direction values greater than a user-defined threshold to 0, and $A$ and $B$ are converted into LAB color space for more accurate color predictions. This results in our final heatmap-weighed mean squared error loss function:

\begin{equation}
\mathcal{L} = H \cdot |A - B|^2_2
\end{equation}

where $A$ is a rendered image from the mesh and $B$ is its corresponding style transferred image acquired during the data preparation phase. 

\begin{figure}
\includegraphics[width=\textwidth]{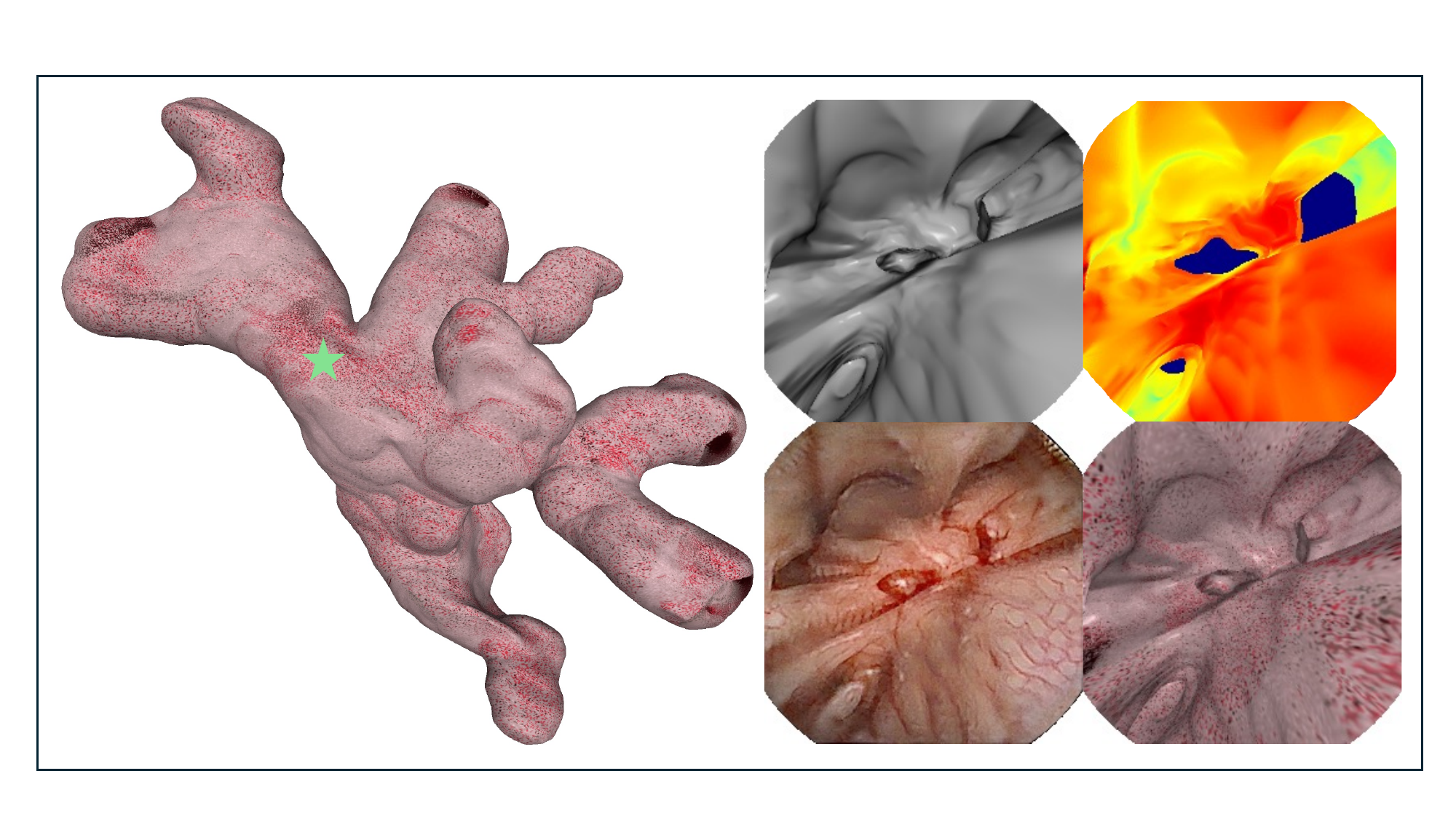}
\caption{Left: visualization of \textit{MeshBrush's} paintings onto the mesh. Right: original renderings, heatmap (see Loss function description), ground-truth style transferred renderings, and renderings from CT using PyTorch3D (from top-left to bottom-right) where the camera is approximately placed at the green star. Corners of the images are masked identically to original endoscopy images.} \label{qualitative}
\end{figure}

\section{Experiments and Results}
We start with a brief overview of our experimental setup. The MLP contains 6 hidden layers each with ReLU activation and a hyperbolic tangent as its final layer. For style transfer, we use Pfeiffer et al.'s laparoscopic image translation module \cite{pfeiffer2019generating} with an in-house patient kidney stone endoscopic surgery dataset. We mask the heatmap in the areas where the distance exceeds 15\,mm, and sample 25 camera poses from the skeletonization for training. PyTorch3D was used as the differentiable rendering library. We evaluated our method on a renal collecting duct mesh, which was manually segmented from a CT scan. The dimensions of our rendered images are (256, 256). To produce the results of this study, we trained our model on an NVIDIA RTX4090 for 300 epochs ($\sim$6 hours). Consistent style transfer outputs are collected by rendering video trajectories from the stylized mesh from the same skeletonization method \cite{skeletor}. 

Qualitative results are displayed in Fig.~\ref{qualitative}, showing various renders of our stylized mesh with corresponding heatmaps for visualization. The quantitative results are shown in Table~\ref{quant}, evaluated on five continuous camera trajectories not in the training set, where we compare the realism (Fréchet Inception Distance (FID)~\cite{heusel2017gans} and Kernel Inception Distance (KID)~\cite{sutherland2018demystifying}) and ORB feature matching capabilities between the untextured, I2I style transferred, and textured images. We use Pyrender for increased realistic renders for the realism metrics. Feature matches are determined to be correct if they correspond to the same 3D point ($< 1$mm). PyTorch3D renders were used for feature matching since they directly provide 3D vertex information for each pixel. Finally, we run sparse reconstruction with COLMAP~\cite{schoenberger2016sfm} on untextured, frame-by-frame I2I, and our outputs (rendered from PyTorch3D) to show that consistent style transfer via mesh stylization is a valid approach for downstream vision-based 3D reconstruction tasks in Fig.~\ref{colmap}. 


\begin{table}[]
\centering
\caption{Quantitative Results of our method. FID~\cite{heusel2017gans} and KID~\cite{sutherland2018demystifying} are computed with respect to the in-house real endoscopy image dataset with a known method~\cite{parmar2021cleanfid}. We report the accuracy of feature matching (the percentage of correctly matched features) and the total number of correct matches, displayed in parenthesis. Untextured and I2I refer to renders from the mesh without any type of style transfer and independently processed style transfers respectively. The bottom two rows display our ablation studies; No HM and No HM \& Fourier represents the evaluation of our model without the heatmap and without both components respectively. The best and second best values for each category is bolded and underlined respectively.}
\begin{tabular}{p{2.5cm} | P{1.5cm}P{1.5cm} | P{2cm}P{2cm}P{2cm}}
\toprule
Method &  FID $(\downarrow)$ & KID $(\downarrow)$ & ORB-1 $(\uparrow)$& ORB-5$(\uparrow)$ & ORB-10$(\uparrow)$\\
\toprule
Untextured & 240.9 & 0.264 & 89.4 (100.9) & \textbf{80.6} (65.8) & \textbf{68.9} (44.8)  \\
I2I & \textbf{106.8} & \textbf{0.089} & 87.5 (164.4) & 66.9 (\underline{94.3}) & 46.0 (\textbf{56.3}) \\
Ours & 206.1 & 0.232 & \textbf{92.1} (\textbf{195.1}) & 66.1 (\textbf{97.2}) & 36.1 (\underline{46.1}) \\
\hline 
No HM & \underline{187.1} & \underline{0.197} & \underline{91.0} (\underline{177.5}) & 64.3 (84.6) & 36.8 (41.6)\\ 
No HM \& Fourier & 246.5 & 0.281 & 87.4 (66.6) & \underline{76.4} (42.3) &  \underline{59.9} (26.7)\\
\toprule
\end{tabular} \label{quant}
\end{table}



\section{Discussion and Conclusion}
The core idea of our work is that stylization \textit{guarantees} consistency throughout the entire mesh; furthermore, it generalizes to any viewpoint within the mesh and does not require any additional training for novel views, which is not the case for traditional consistent style transfer methods.

\textbf{Realism.} The FID and KID scores measure the distance from real endoscopy images in terms of fidelity. The untextured renderings have the lowest realism, while the style transfer (I2I) outputs have the highest. Because the I2I images supervise our mesh texturing, it is the upper bound for realism. We note that the realism of the output images is limited since we rely on physics-based renderings rather than direct optimization such as in generative models with discriminator networks. However, we observe specific improvements over rendering from non-textured meshes. In addition, we found that FID and KID relied heavily on our rendering strategy because we stylized the mesh instead of the images directly. 
Indeed, we found that rendering from Pyrender instead of PyTorch3D dramatically changes the FID and KID scores. 

For the ablation study, we observe that once we remove the heatmap from the loss function, the realism metrics improve whereas feature matching deteriorates. Ablating both components prevented the network from learning any high-frequency information, causing the mesh to be mostly monochrome. Visually, removing the heatmap decreased the color variation of the mesh, only allowing the model to learn pink to gray colors. 

\textbf{Feature Matching.} Because our method renders from a stylized mesh, features are expected to persist accurately between frames. For ORB-1, i.e. neighboring frames, our method outperforms all other baselines in both the number of features correctly matched and the percentage of correctly matched features. In addition, we show the effectiveness of the heatmap-based loss and positional encodings in robust feature matching. However, the performance of \textit{MeshBrush} deteriorates for ORB-5 and ORB-10. This may be attributed to the speckle-like pattern that is often generated by our model. Given the fact that cameras are also extremely close to the anatomical surface, these patterns are often sharp and excessive. The abundance of speckles may reduce the performance of feature matching, decreasing the overall performance in short sequences.  Although our model seems to deteriorate across short-term sequences, we demonstrate that the detected features are unique and robust enough for \textit{global} matching. 

\begin{figure}
\centering 
\includegraphics[width=\textwidth]{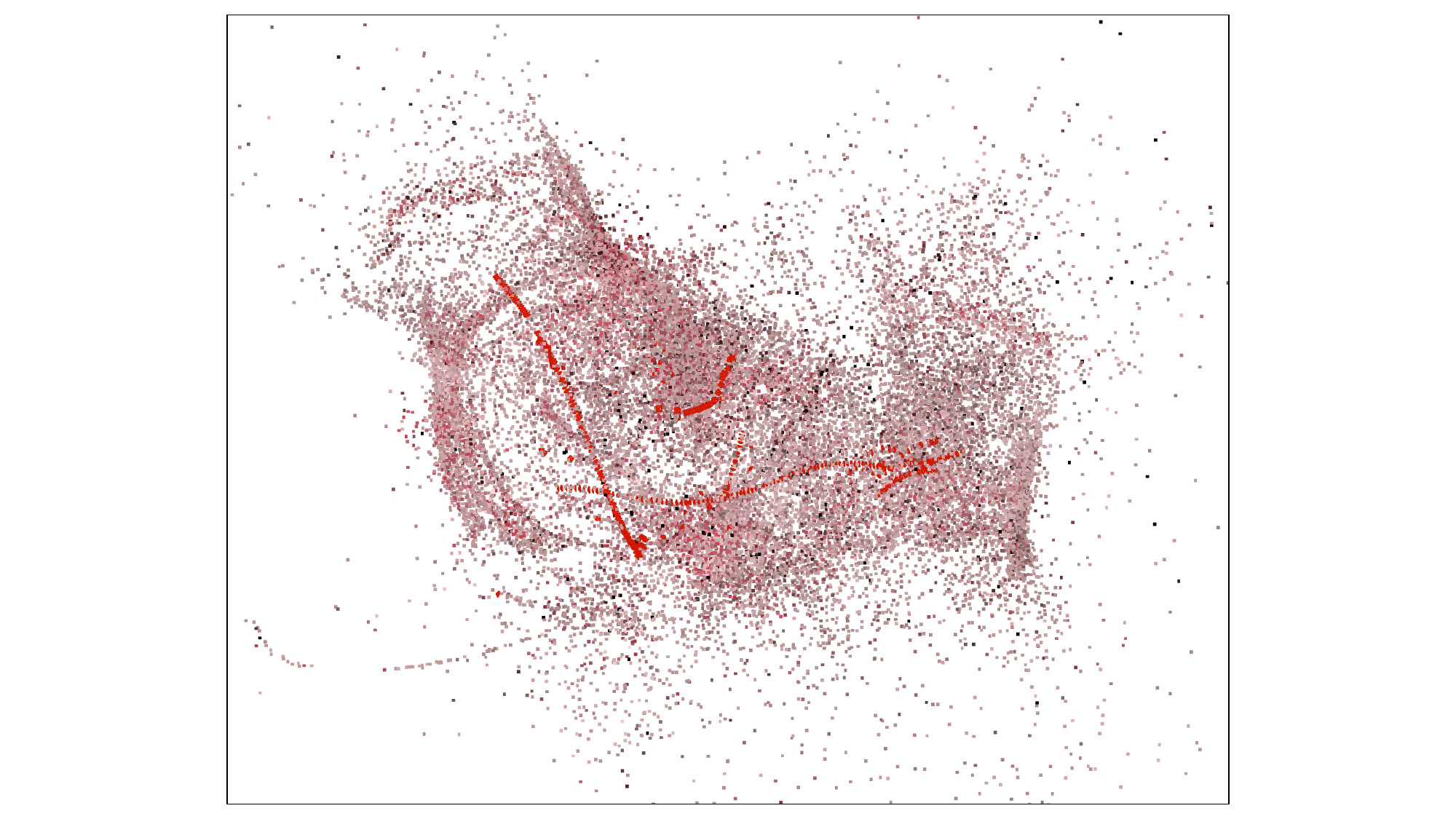}
\caption{Sparse COLMAP reconstruction using renders from the stylized mesh. Red squares show the estimated camera trajectory} \label{colmap}
\end{figure}

\textbf{Structure-from-Motion.} The sparse reconstructions of COLMAP, for both PyTorch3D and Pyrender renderings of the mesh, failed for both untextured and I2I methods. This confirms that textures from untextured and I2I-translated images do not provide unique descriptors for global camera localization. On the other hand, the features predicted by \textit{MeshBrush} have globally distinct descriptors, shown in Fig.~\ref{colmap}. We found that COLMAP uses 41.3\% from 1,120 images to reconstruct 32327 points. Previous work achieved an average of 20.6\% from 498 images to reconstruct 4838 points for kidney reconstructions from rendered views~\cite{acar2023towards}. Thus, our method ensures the global and long-term consistency necessary for sophisticated downstream vision tasks like 3D reconstruction. 

We present a mesh stylization technique for endoscopy to synthesize consistent style transfer views. Our method leverages existing image-to-image style transfer networks to create textures on the mesh to imitate real patient anatomy. While this method has only been tested for a renal collecting duct model, the proposed method is generic and self-supervised, thus has the potential to translate to other anatomies and procedures. 

\begin{credits}
\subsubsection{\ackname} This study was partially supported by the NIH T32 Training Grant (grant number T32EB021937) and the Vanderbilt Institute of Surgery and Engineering Seed Grant. 
\subsubsection{\discintname}
The authors have no competing interests to declare that are
relevant to the content of this article.
\end{credits}
\bibliographystyle{splncs04}
\bibliography{Paper-2259}

\begin{thebibliography}{10}
\providecommand{\url}[1]{\texttt{#1}}
\providecommand{\urlprefix}{URL }
\providecommand{\doi}[1]{https://doi.org/#1}

\bibitem{acar2023towards}
Acar, A., Lu, D., Wu, Y., Oguz, I., Kavoussi, N., Wu, J.Y.: Towards navigation in endoscopic kidney surgery based on preoperative imaging. Healthcare Technology Letters  (2023)

\bibitem{skeletor}
Au, O., Tai, C.L., Chu, H.K., Cohen-Or, D., Lee, T.Y.: Skeleton extraction by mesh contraction. ACM Trans. Graph.  \textbf{27} (08 2008)

\bibitem{Canfes_2023_WACV}
Canfes, Z., Atasoy, M.F., Dirik, A., Yanardag, P.: Text and image guided 3d avatar generation and manipulation. In: Proceedings of the IEEE/CVF Winter Conference on Applications of Computer Vision (WACV). pp. 4421--4431 (January 2023)

\bibitem{cardoso2023exploring}
Cardoso, S.A., Suyambu, J., Iqbal, J., Jaimes, D.C.C., Amin, A., Sikto, J.T., Valderrama, M., Aulakh, S.S., Ramana, V., Shaukat, B., et~al.: Exploring the role of simulation training in improving surgical skills among residents: A narrative review. Cureus  \textbf{15}(9) (2023)

\bibitem{chu2023video}
Chu, E., Lin, S.Y., Chen, J.C.: Video controlnet: Towards temporally consistent synthetic-to-real video translation using conditional image diffusion models. arXiv preprint arXiv:2305.19193  (2023)

\bibitem{engelhardt2018improving}
Engelhardt, S., De~Simone, R., Full, P.M., Karck, M., Wolf, I.: Improving surgical training phantoms by hyperrealism: deep unpaired image-to-image translation from real surgeries. In: Medical Image Computing and Computer Assisted Intervention. pp. 747--755. Springer (2018)

\bibitem{frischtemporal}
Frisch, Y., Fuchs, M., Mukhopadhyay, A.: Temporally consistent sequence-to-sequence translation of cataract surgeries. In: International Journal of Computer Assisted Radiology and Surgery. pp. 1217--1224. IJCARS (2023)

\bibitem{frisch2023temporally}
Frisch, Y., Fuchs, M., Mukhopadhyay, A.: Temporally consistent sequence-to-sequence translation of cataract surgeries. International Journal of Computer Assisted Radiology and Surgery pp.~1--8 (2023)

\bibitem{funke2018generative}
Funke, I., Bodenstedt, S., Riediger, C., Weitz, J., Speidel, S.: Generative adversarial networks for specular highlight removal in endoscopic images. In: Medical Imaging: Image-Guided Procedures, Robotic Interventions, and Modeling. SPIE (2018)

\bibitem{heusel2017gans}
Heusel, M., Ramsauer, H., Unterthiner, T., Nessler, B., Hochreiter, S.: Gans trained by a two time-scale update rule converge to a local nash equilibrium. Advances in neural information processing systems  \textbf{30} (2017)

\bibitem{Hollein_2022_CVPR}
H\"ollein, L., Johnson, J., Nie{\ss}ner, M.: Stylemesh: Style transfer for indoor 3d scene reconstructions. In: Proceedings of the IEEE/CVF Conference on Computer Vision and Pattern Recognition (CVPR). pp. 6198--6208 (June 2022)

\bibitem{KANG2023101198}
Kang, H., Dong, X., Cao, J., Chen, Z.: Neural style transfer for 3d meshes. Graphical Models  \textbf{129},  101198 (2023)

\bibitem{khan2019sim}
Khan, R., Scaffidi, M., Grover, S., Gimpaya, N., Walsh, C.: Simulation in endoscopy: Practical educational strategies to improve learning. World Journal of Gastroenterology  \textbf{11}(3),  209--218 (2019)

\bibitem{khan2018virtual}
Khan, R., Plahouras, J., Johnston, B.C., Scaffidi, M.A., Grover, S.C., Walsh, C.M.: Virtual reality simulation training for health professions trainees in gastrointestinal endoscopy. Cochrane Database of Systematic Reviews (8) (2018)

\bibitem{liang2023flowvid}
Liang, F., Wu, B., Wang, J., Yu, L., Li, K., Zhao, Y., Misra, I., Huang, J.B., Zhang, P., Vajda, P., et~al.: Flowvid: Taming imperfect optical flows for consistent video-to-video synthesis. arXiv preprint arXiv:2312.17681  (2023)

\bibitem{lu2023assist}
Lu, D., Wu, Y., Acar, A., Yao, X., Wu, J.Y., Kavoussi, N., Oguz, I.: Assist-u: A system for segmentation and image style transfer for ureteroscopy. Healthcare Technology Letters  (2023)

\bibitem{ma2023xmesh}
Ma, Y., Zhang, X., Sun, X., Ji, J., Wang, H., Jiang, G., Zhuang, W., Ji, R.: X-mesh: Towards fast and accurate text-driven 3d stylization via dynamic textual guidance. In: Proceedings of the IEEE/CVF International Conference on Computer Vision. pp. 2749--2760 (2023)

\bibitem{parmar2021cleanfid}
Parmar, G., Zhang, R., Zhu, J.Y.: On aliased resizing and surprising subtleties in gan evaluation. In: CVPR (2022)

\bibitem{pfeiffer2019generating}
Pfeiffer, M., Funke, I., Robu, M.R., Bodenstedt, S., Strenger, L., Engelhardt, S., Ro{\ss}, T., Clarkson, M.J., Gurusamy, K., Davidson, B.R., et~al.: Generating large labeled data sets for laparoscopic image processing tasks using unpaired image-to-image translation. In: Medical Image Computing and Computer Assisted Intervention--MICCAI. pp. 119--127. Springer (2019)

\bibitem{rau2019implicit}
Rau, A., Edwards, P.E., Ahmad, O.F., Riordan, P., Janatka, M., Lovat, L.B., Stoyanov, D.: Implicit domain adaptation with conditional generative adversarial networks for depth prediction in endoscopy. International journal of computer assisted radiology and surgery  \textbf{14},  1167--1176 (2019)

\bibitem{rivoir2021longterm}
Rivoir, D., Pfeiffer, M., Docea, R., Kolbinger, F., Riediger, C., Weitz, J., Speidel, S.: Long-term temporally consistent unpaired video translation from simulated surgical 3d data (2021)

\bibitem{schoenberger2016sfm}
Sch\"{o}nberger, J.L., Frahm, J.M.: Structure-from-motion revisited. In: Conference on Computer Vision and Pattern Recognition (CVPR) (2016)

\bibitem{Sharan_2022}
Sharan, L., Romano, G., Koehler, S., Kelm, H., Karck, M., De~Simone, R., Engelhardt, S.: Mutually improved endoscopic image synthesis and landmark detection in unpaired image-to-image translation. Journal of Biomedical and Health Informatics  \textbf{26}(1),  127–138 (Jan 2022)

\bibitem{SHINOMIYA201813}
Shinomiya, A., Shindo, A., Kawanishi, M., Miyake, K., Nakamura, T., Matsubara, S., Tamiya, T.: Usefulness of the 3d virtual visualization surgical planning simulation and 3d model for endoscopic endonasal transsphenoidal surgery of pituitary adenoma: Technical report and review of literature. Interdisciplinary Neurosurgery  \textbf{13},  13--19 (2018)

\bibitem{sutherland2018demystifying}
Sutherland, J., Arbel, M., Gretton, A.: Demystifying mmd gans. In: International Conference for Learning Representations. pp. 1--36 (2018)

\bibitem{tancik2020fourier}
Tancik, M., Srinivasan, P., Mildenhall, B., Fridovich-Keil, S., Raghavan, N., Singhal, U., Ramamoorthi, R., Barron, J., Ng, R.: Fourier features let networks learn high frequency functions in low dimensional domains. Advances in Neural Information Processing Systems  \textbf{33},  7537--7547 (2020)

\bibitem{teed2020raft}
Teed, Z., Deng, J.: Raft: Recurrent all-pairs field transforms for optical flow. In: Computer Vision--ECCV. pp. 402--419. Springer (2020)

\bibitem{st_depth_tong}
Tong, H.S., Ng, Y.L., Liu, Z., Ho, J., Chan, P.L., Chan, J., Kwok, K.W.: Real-to-virtual domain transfer-based depth estimation for real-time 3d annotation in transnasal surgery: a study of annotation accuracy and stability. International Journal of Computer Assisted Radiology and Surgery  \textbf{16},  731--739 (2021)

\bibitem{venkatesh2020unsupervised}
Venkatesh, V., Sharma, N., Srivastava, V., Singh, M.: Unsupervised smoke to desmoked laparoscopic surgery images using contrast driven cyclic-desmokegan. Computers in Biology and Medicine  \textbf{123},  103873 (2020)

\bibitem{zhang2023adding}
Zhang, L., Rao, A., Agrawala, M.: Adding conditional control to text-to-image diffusion models. In: Proceedings of the IEEE/CVF International Conference on Computer Vision. pp. 3836--3847 (2023)

\bibitem{zhu2017unpaired}
Zhu, J.Y., Park, T., Isola, P., Efros, A.A.: Unpaired image-to-image translation using cycle-consistent adversarial networks. In: Proceedings of the IEEE international conference on computer vision. pp. 2223--2232 (2017)

\end{thebibliography}
%




\end{document}